\documentclass[12pt]{iopart}

\usepackage{graphicx}
\usepackage{subfigure}
\usepackage{caption}
\usepackage{cite}
\usepackage[numbers,sort&compress,comma]{natbib} 
\usepackage[hidelinks]{hyperref} 
 
\begin{document}

\vspace*{-2cm} 
\title[ ]{Quantum-Optically Resolving the Number of Colloidal Quantum Dots in a Subwavelength Volume}

\author{
Zhi-Bo Ni$^{1,2}$,
Jia-Wang Yu$^{1,2}$,
Jiong-Zhao Li$^{3}$,
Xiao-Tian Cheng$^{1,2}$,
Mei-Na Jiang$^{3}$,
Zi-Xuan Song$^{1,2}$,
Xiao-Qing Zhou$^{4}$,
Wei Fang$^{5}$,
Chen-Hui Li$^{1,2,6}$,
Feng Liu$^{1}$,
Xing Lin$^{1,2,\ast}$,
Chao-Yuan Jin$^{1,2,7,\dagger}$
}

\address{$^{1}$ State Key Laboratory of Silicon and Advanced Semiconductor Materials \& College of Information Science and Electronic Engineering, Zhejiang University, Hangzhou 310027, China}
\address{$^{2}$ ZJU-Hangzhou Global Scientific and Technological Innovation Center, Zhejiang University, Hangzhou 311200, China}
\address{$^{3}$ Department of Chemistry, Zhejiang University, Hangzhou 310027, China}
\address{$^{4}$ Department of Physics, School of Science, Westlake University, Hangzhou 310030, China}
\address{$^{5}$ College of Optical Science and Engineering, Zhejiang University, Hangzhou 310027, China}
\address{$^{6}$ Research Institute of Intelligent Computing, Zhejiang Lab, Hangzhou 311121, China}
\address{$^{7}$ College of Integrated Circuits, Zhejiang University, Hangzhou 311200, China}

\ead{$^\ast$lxing@zju.edu.cn; $^\dagger$jincy@zju.edu.cn} 

\vspace{10pt}
\begin{indented}
\item[]
\end{indented}

\begin{abstract}
The number resolution of solid-state artificial atoms is of fundamental interest for the study of quantum few-body systems, yet remains experimentally challenging. Quantum optical experiments offer a non-invasive approach which links up macroscopic measurements with the quantity of quantum emitters. In this work, we propose a time-domain quantum optical methodology for the strict numbering of colloidal CdSe/CdS/ZnS quantum dots (QDs) confined in subwavelength-size polystyrene capsules. The non-polarized, homogeneously broadened emission of colloidal QDs in the subwavelength volume satisfies the description of Dicke's superradiance of identical quantum emitters. An analytic relation describes the numerical dependence of the second-order photon correlation on the number and the collective lifetime of emitters, yielding an experimental counting range of colloidal QDs from one to ten. This work provides a robust pathway for the non-invasive numbering of artificial atoms and the investigation of collective light-matter interactions at the nanoscale.
\end{abstract}

\clearpage

\section{Introduction}

Quantum few-body systems based on neutral atoms\cite{Will2010,Goban2018,goban2015superradiance,zaccanti2009observation,isenhower2010demonstration}, ions\cite{monroe2013scaling}, photons\cite{cao2024multi,kielinski2024ghz,douglas2016photon}, electrons\cite{shaju2025evidence}, quasi-particles\cite{sun2017direct}, or artificial atoms\cite{grim2019scalable,Scheibner2007,bradac2017room,angerer2018superradiant} have attracted extensive interest for the study of fundamental physics and the development of quantum information technologies. After the discovery of macroscopic quantum phenomena such as superconductivity\cite{yanase2003theory}, quantum Hall effect \cite{ghazaryan2017light}, Bose-Einstein condensation\cite{griffin1996bose,giorgini2008theory} and quantum fluids of light\cite{carusotto2013quantum}, the accurate manipulation of quantum few-body systems is now leading to the development of quantum entangled states\cite{cao2024multi,kielinski2024ghz,ballance2015hybrid,nadlinger2022experimental} and cluster states \cite{lib2024resource,ferreira2024deterministic}, which are the recent focus of scalable quantum computing.

When harnessing the quantum nature of few-body systems continuously advances, resolving the number of interacting particles at microscopic levels remains a challenging and non-trivial task. Pioneered by Millikan's oil drop experiment \cite{millikan1913elementary, millikan1924electron}, macroscopic electrical measurement has been set as the main approach to count the number of electrons\cite{shaju2025evidence} or photons\cite{divochiy2008superconducting}. In parallel, electronic microscopes have been employed as a microscopic tool to inspect artificial atoms \cite{lin2021analytical, wang2020formation}. Although representing the ultimate technology for the study of materials at atomic levels\cite{kita2007multidirectional}, 3D tomography of artificial structures is usually destructive and cost-ineffective , which is not suitable for non-invasive numbering of artificial atoms. 

Very recently, collective spontaneous emission known as superradiance has been observed in nitrogen vacancy (NV) centers in diamond nanocrystals \cite{angerer2018superradiant, bradac2017room, pallmann2024cavity}. A notable correlation between the decay time of spontaneous emission and the number of color centers has been reported \cite{bradac2017room}. This nano-diamond system holds a few advantages for the observation of collective quantum behaviors, such as narrow emission linewidth, high brightness, and subwavelength confinement for emitters. However, since the identicality of the artificial atoms is obscured by the random distribution of the emission wavelength as well as the polarization orientation of the NV centers \cite{christinck2020characterization, pallmann2024cavity}, this approach has yet to be able to accurately determine the number of NV centers. The numerical relation between radiative emission and the number of NV centers is highly dependent on the selection of experimental datasets \cite{bradac2017room}. Nevertheless, those preliminary experiments indicate that one could possibly think about strictly resolving the number of artificial atoms from macroscopic measurements similar to those electrical experiments on electrons or photons \cite{shaju2025evidence, divochiy2008superconducting}, as long as a delicate calibration methodology could be established within a quantum optical framework for solid-state artificial atoms.

Here, we demonstrate a number-resolving technology based on macroscopic measurement of CdSe/CdS/ZnS quantum dots (QDs) that are confined by polystyrene capsules within a subwavelength volume. The second-order photon correlation function and the collective spontaneous emission lifetime of the QD ensemble are utilized as quantum optical tools for macroscopic characterization. The core-shelled QDs possess a non-polarized dipole orientation and a homogeneously broadened emission peak, exhibiting very different nature of radiative emission compared to NV centers in nano-diamonds\cite{chung2003room,christinck2020characterization}. Right because of the homogeneously broadened emission of non-polarized single dots, high transition uniformity among different QDs can be readily expected. In addition, subwavelength confinement within a few tens of nanometers is provided by polystyrene capsules for nanocrystals, which satisfies the subwavelength condition of the quantum Dicke model. Therefor, a new methodology based on the weakly coupled emitter system is built up to simplify the analytic description of the collective quantum behavior through spontaneous supperradiance, which sets a brand new ground for the macroscopic non-invasive inspection of quantum few-body systems. 

\section{Results and Discussion}
\subsection{Number-Resolving Methodology for Quantum Emitters}

In the description of superradiance, the radiative decay rate of an ensemble of quantum emitters is enhanced with a fluorescence lifetime inversely scaling with the number of emitters, $\tau_c \propto 1/N$. According to the quantum Dicke model, such scaling law is valid only under specific conditions that all emitters couple identically to a common radiation mode and experience uniform mutual coupling strengths. In the ensemble of solid-state emitters, these coupling strengths vary due to spatial and spectral randomness. Consequently, it is hard to apply the quantum Dicke model to a given ensemble to precisely predict the number of emitters.

Colloidal quantum dots (CQDs) exhibit a simple excitonic structure that is approximated by a two-level system, closely resembling the energy configuration of atomic emitters. Owing to this similarity, the quantum optical model developed for two-level quantum emitters provides an appropriate theoretical foundation for CQDs \cite{masson2022universality,patti2021controlling}. A framework of the number-resolving process is derived from the quantum Dicke model, which describes the coupling between a quantized field and an ensemble of two-level systems. Within a subwavelength volume, $N$ emitters are randomly distributed, which can be described by the following Hamiltonian and Lindblad dissipator
\begin{equation}
H = \hbar \sum_{i} \omega_0 \sigma_i^{\dagger} \sigma_i + \hbar \sum_{i \ne j} J_{ij} \sigma_i^{\dagger} \sigma_j,
\end{equation}
\begin{equation}
	\mathcal{D}[\rho] = \sum_{i,j=1}^{N} \frac{\Gamma_{ij}}{2} \left( 2\sigma_j \rho \sigma_i^\dagger - \sigma_i^\dagger \sigma_j \rho - \rho \sigma_i^\dagger \sigma_j \right).
\end{equation}
Here, $\rho$ denotes the atomic density matrix, and 
$\sigma_i = |g_i\rangle \langle e_i|$ is the lowering operator of the $i$-th atom. 
$J_{ij}$ represents the coherent coupling strength between atoms $i$ and $j$, 
describing the lossless exchange of energy mediated by virtual photons. 
In contrast, $\Gamma_{ij}$ characterizes the collective decay rate, accounting for irreversible radiative losses that drive dissipative and collective phenomena such as superradiance and subradiance \cite{patti2021controlling,gonzalez2024light}.

For multiple quantum emitters confined in a subwavelength volume, the radiative decay rate of the ensemble is modified compared to that of isolated emitters\cite{Scheibner2007}. As the number of coupled emitters increases, the antibunching feature in photon statistics gradually vanishes. This behavior is interpreted as the emergence of collective behavior among emitters \cite{masson2022universality}.

\begin{equation}
	g^{(2)}(0)= 1+\frac{1}{N} \left[\rm{Var} \left( \frac{\{\Gamma_\nu\}}{\bar{\Gamma}_0} \right)-1\right] - \frac{2}{N} \,\mathrm{Var}\!\left( \frac{\Gamma_0^{i}}{\bar{\Gamma}_0} \right),
\end{equation}
where $\bar{\Gamma}_0=1/\bar{\tau}_0$ denotes the average photon decay rate of isolated emitters, $\Gamma_0^{i}$ represents the photon decay rate of the $i$-th emitter, and $\Gamma_v$ corresponds to the decay rate of the $v$-th photon decay channel\cite{reitz2022cooperative}. The third term in the equation originates from the inhomogeneity among quantum emitters, which introduces deviations from the idealized uniform emitter scenario. 

Remarkable progress has been made in the synthesis technique of CQDs, leading to enhanced stability through ligand engineering and core-shelled structures\cite{peng1997epitaxial}. Precise control over the size and emission wavelength of CQDs, ranging from individual nanocrystals to stoichiometrical ensembles, has been achieved elsewhere \cite{peng2000shape,Peng2001,huang2021synthesis}. These highly efficient and nearly identical emitters effectively reduce the adversely decoupling effects caused by inhomogeneous broadening. Despite broad emission spectra, CQDs exhibit a nearly consistent emission peak in comparison with other artificial atoms, leading to notable spectral overlap among QDs. As a result, individual emitters within CQD ensembles can be treated as identical. Upon this assumption, the expression for $g^{(2)}(0)$ reduces to
\begin{equation}
	g^{(2)}(0)= 1+ \frac{1}{N}\left[\frac{(N - 1)\Gamma_c^2}{N^2 \bar{\Gamma}_0^2}-1\right].
\end{equation}
Here, $\Gamma_c$ ($\Gamma_c=1/\tau_c$) represents the collective radiative decay rate of the CQD ensemble, quantifying the cooperative enhancement of emission arising from inter-emitter coupling. This simple form highlights the dependence of the second-order correlation at zero-time delay on the number and collective decay rate of quamtum emitters relative to individual ones. A detailed derivation and discussion of the underlying assumptions are provided in \textit{Methods}. Based on the above equation, the number of CQDs is determined by the values of the second-order correlation at zero-time delay and the collective decay rate. Both values are from macroscopic measures of the CQD ensemble, which are fully based on optical features of quantum few-body systems. It thus provides a self-consistent estimation of $N$ that remains reliable under varying inter-emitter coupling strengths, enabling accurate number extraction across a broad range of coupling conditions.

In the previous study of superradiance in artificial atoms, the relation between emitter number $N$ and collective lifetime $\tau_c$ is widely reported, as collective fluorescence lifetime decreases in proportion to the number of quantum emitters \cite{bradac2017room, goban2015superradiance}. This relationship is often modeled by the expression $\tau_c = a/N$, where $a$ is a constant related to the inter-emitter coupling strength. By measuring the collective lifetime $\tau_c$, the emitter number $N$ is then inferred using this scaling law. However, this method only represents the statistical fitting with a free parameter $a$, rather than a deterministic number resolving. In statistical fitting, the reliance on the inter-emitter coupling strength makes the value of $N$ highly sensitive to the selection of the dataset as well as the calibration method \cite{bradac2017room}. Alternatively, another approach based on the relation between second-order correlation at zero-time delay $g^{(2)}(0)$ and the number of quantum emitters $N$ is also applied in a limited scope. For conditions that quantum emitters are uncoupled, $g^{(2)}(0)=1-1/N$ is derived as the scaling law for $g^{(2)}(0)$ depending on the emitter number $N$\cite{grim2019scalable,meijer2005generation}. However, when the inter-emitter coupling starts to emerge, this description for $g^{(2)}(0)$ is invalid.

In contrast to those conventional methods, our approach using equation (4) inherently captures the influence of collective emission, making the analysis valid even when the inter-emitter coupling strength is varied. As a result, it provides a robust and consistent estimation of the emitter number in homogeneously broadened and non-polarized CQD ensembles, extending the applicability of the number-dependent quantum optical measurements into coupled-emitter regime.

\subsection{Experimental Design and Preparatory Steps}

Figure 1 illustrates the experimental framework for the number resolution of CQDs. CQDs are confined within a subwavelength volume. The quantum emitters are optically indistinguishable both spatially and spectrally. Time-resolved photoluminescence (TRPL) and photon-correlation measurements record the arrival time of single photons, providing direct access to instantaneous emission dynamics \cite{fox2006quantum}. A time-domain micro-photoluminescence setup is depicted in Figure 1(b) by setting up a Hanbury Brown and Twiss (HBT) measurement using two single-photon avalanche diodes (SPADs). In Figure 1(c), when multiple QDs are confined within a subwavlength volume, their collective emission leads to faster photon decay compared to that of isolated single emitters \cite{Scheibner2007}. The antibunching feature of the photon statistics is expected to disappear with a large $N$ \cite{kim2018super, grim2019scalable}. The emitter number $N$ is expressed as a function of the characteristic parameters $\bar{\Gamma}_0$, $\Gamma_c$, and $g^{(2)}(0)$, following the analytic relation summarized in equation (4).

In Dicke's model, superradiance arises when $N$ identical oscillators interact coherently through a common optical field. The collective mode constructs an average decay rate $\Gamma_{c}$ by diagonalizing the loss matrix $\Gamma_{ij}$. Homogeneously broadened CQDs create nearly ideal conditions for uniform optical coupling among multiple emitters, making them possibly the closest analogs of the quantum few-body system governed by Dicke's model, compared to other solid-state systems such as epitaxial QDs or NV centers. In addition, the substrate-free nature of CQDs enables the controllable positioning of artificial quantum emitters \cite{abramson2012quantum, manfrinato2013controlled, xie2015nanoscale, makey2020universality}. Core-shelled CdSe/8CdS/9ZnS QDs are employed and confined by polystyrene capsules. The exciton Bohr radius of three-dimensional confined CdSe QDs was calculated as 4.9 nm based on material parameters\cite{sun2008linear}. Our core-shelled structures with an out-shell diameter of around 15.7 nm effectively suppresses the direct charge transfer from dot to dot, while the polystyrene capsule with a diameter of around 60 nm ensures that CQDs are closely packed within a subwavelength volume, guaranteeing effective and uniform optical coupling among CQDs. 

\begin{center}
  \includegraphics[width=0.9\textwidth]{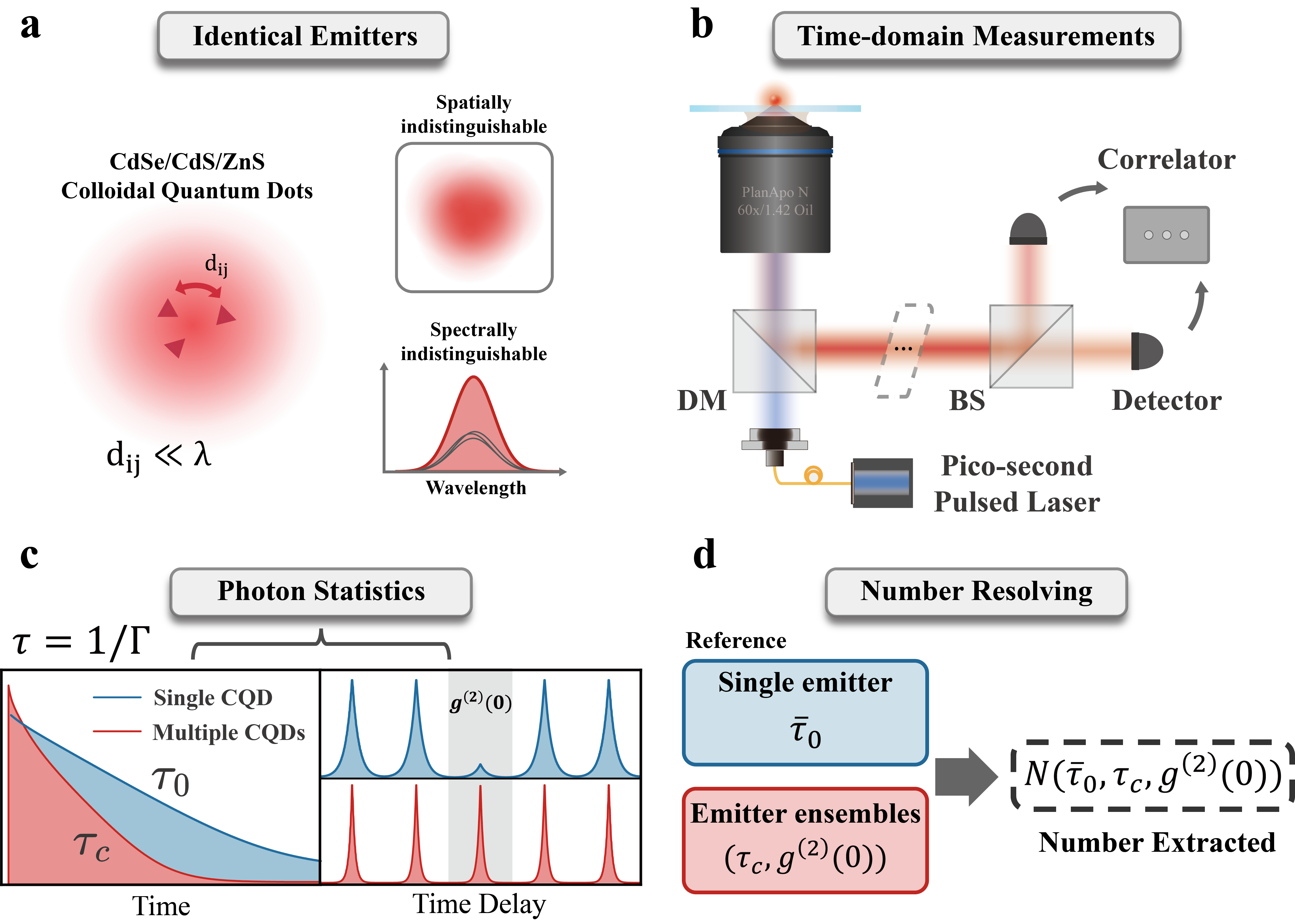}
\end{center}
\vspace{0pt}
\noindent \textbf{Figure 1: Experimental framework for the number resolution of spatially confined colloidal quantum dots (CQDs).} (a) Illustration of multiple CQDs confined within a subwavelength volume, where the inter-dot spacing is much smaller than the emission wavelength. The emitters are almost identical and cannot be distinguished either spatially or spectrally by optical means. (b) Schematic of the micro-photoluminescence system, with emphasis on the time-correlated single photon counting measurement based on a pulsed laser and two single photon detectors. (c) Collective coupling among multiple emitters leads to radiative properties that deviate from the single-emitter response. (d) By comparing the modified radiative dynamics with those of single isolated emitters used as a reference, the number of QDs confined in the subwavelength region can be quantitatively resolved.
\vspace{15pt} 

Preserving the optical performance of CQDs during encapsulation is also a crucial task for reliable optical measurements. Thermal polymerization of organic polymers is employed during the preparation stage \cite{Gao2008, lohmann2022controlling, Ding2004}. These polymers have long been recognized for their negligible toxicity and excellent biocompatibility\cite{yang2004encapsulation, bottrill2011some}; more recently, their utility has been extended to quantum optical applications\cite{Negele2013}. In particular, polystyrene is chosen over inorganic materials such as non-crystal silica because it avoids ligand exchange and offers superior solubility for CQDs, which reduces chemical aggregation and precipitation during synthesis \cite{Selvan2005, Zhang2020, Koole2008, Zhang2008, Acebron2015}. For this reason, styrene is used as the precursor material for thermal polymerization to form the encapsulating shell.

\begin{center}
  \includegraphics[width=\textwidth]{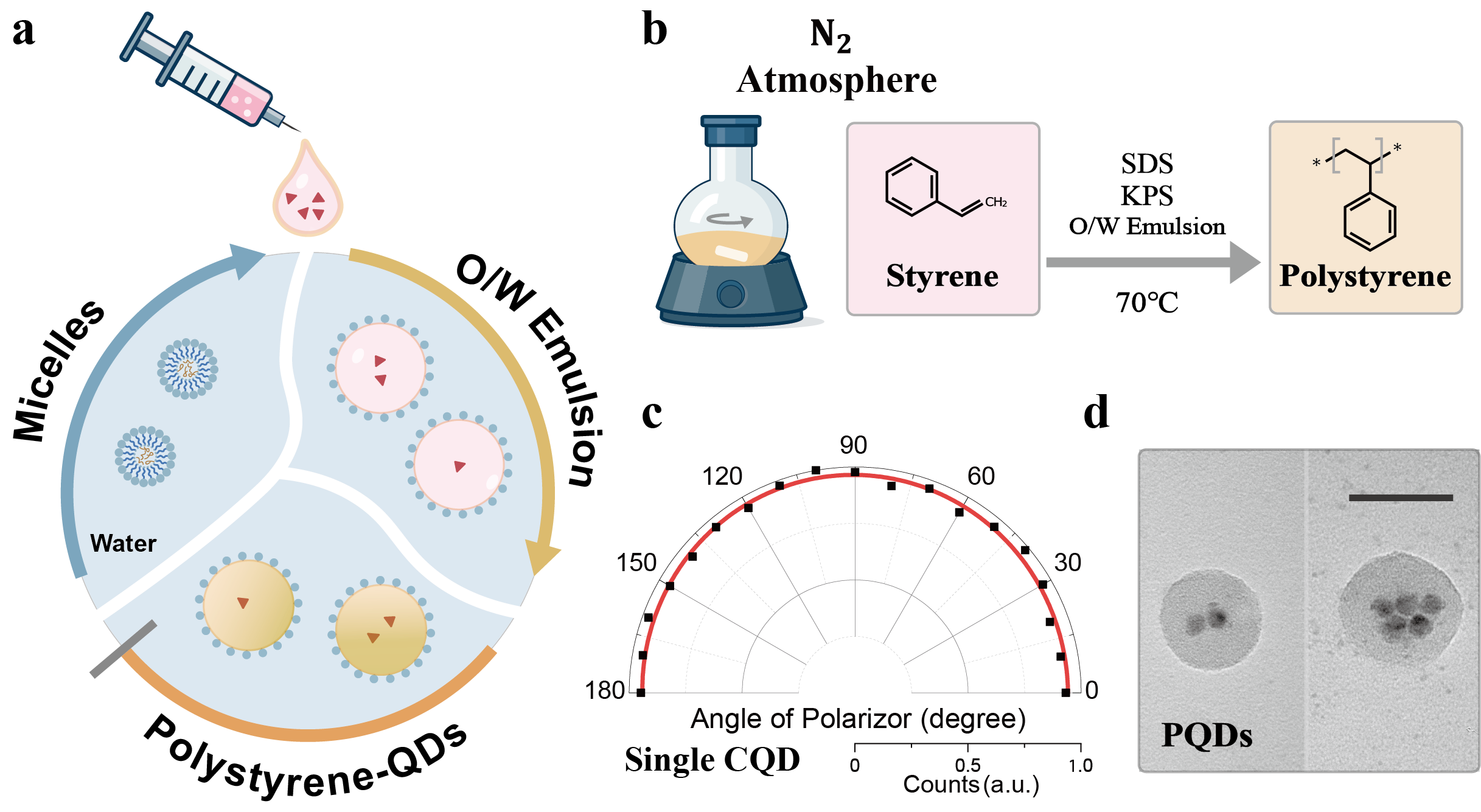} 
\end{center}
\vspace{0pt}
\noindent \textbf{Figure 2: Sketch of the synthesis process of polystyrene encapsulated CQDs.} (a) Styrene undergoes thermal polymerization via an oil-in-water emulsion method. CQDs are pre-dispersed in styrene, achieving polystyrene encapsulation during the process. (b) The thermal polymerization from styrene to polystyrene is performed under a nitrogen-protected atmosphere with vigorous stirring. (c) Polarization-dependent emission of a single CQD encapsulated in a polystyrene nanoparticle, showing an almost non-polarized radiative response. (d) Transmission electron microscopy images of PQDs. Scale bar: 50 nm.
\vspace{15pt} 

Figure 2(a) depicts the synthesis process of polystyrene encapsulated quantum dots (PQDs) \cite{sjoblom1996microemulsions,sheng2006situ}. The process starts when surfactant molecules in water self-assemble into micelles. These micelles provide nanoscale compartments for the subsequent reaction. Styrene monomer containing dispersed CQDs is then introduced into the aqueous phase to form a stable oil-in-water emulsion. In this emulsion, the surfactant controls the droplet size and prevents aggregation between droplets. CQDs are distributed within the oil phase and remain confined in the styrene droplets. Finally, heating triggers thermal polymerization of styrene, transforming the liquid droplets into solid polystyrene nanospheres that encapsulate the CQDs.

Figure 2(d) shows transmission electron microscope (TEM) images of PQDs. The inner shadow projection clearly confirms that CQDs are embedded within individual polystyrene nanoparticles, demonstrating effective encapsulation and spatial confinement. The lower panel in Figure 2(c) presents the polarization-dependent emission of a single CQD, obtained by rotating a linear polarizer in the detection path. The measured intensity remains nearly circular with respect to the polarization angle, yielding a degree of polarization (DOP) of approximately 4.3\%. This weak angular dependence indicates that the emission is nearly polarization-independent, corresponding to a non-polarized dipole orientation that is effectively averaged by the surrounding environment.

\subsection{Optical Characterization of Encapsulated Colloidal Quantum Dots }
Figure 3 summarizes the static and time-resolved photoluminescence (PL) spectra of CQDs and PQDs in different solutions. Fluorescence from the QD ensemble is collected by an air objective with a focal point inside the liquid. As shown in Figure 3(a), the fluorescence spectrum of PQDs in water remains almost the same as that of CQDs in oil-phase solution. The three curves overlap largely with imperceptible distinctions. The inset summarizes the quantum yield (QY) of CQDs in different environments. For PQDs, the scattering caused by polystyrene spheres may introduce some deviation in the results. By averaging multiple samples, the QY of PQDs exceeding 80\% is determined. These results indicate that the optical properties of PQDs are well preserved during the polymer encapsulation process\cite{Negele2013}. Furthermore, the observed lifetime variation of the CQD ensembles in different solutions matches the theoretical prediction. The detailed analysis is provided in \textit{Supplementary Information}. 

For single-particle optical measurements, PQDs were spin-coated onto quartz substrates. $\rm 0.13\mu m$-thick quartz glasses were chosen to optimize optical performance with a 60$\times$ oil-immersed objective. The PQDs were then diluted and re-dispersed in ethanol to achieve a uniform distribution. A home-built micro-photoluminescence system enabled the distinction of single polystyrene nanoparticles encapsulating CQDs. Figure 3(b) compares the TRPL spectra of individual CQDs and PQDs. The lifetime of a single PQD is slightly shorter than that of a bare CQD, which can be attributed to environment variations. Photon correlation measurements were performed using the HBT configuration to analyze the distribution of detection time intervals between two detectors \cite{fox2006quantum}. Filtered emission photons were split by a 50:50 beam splitter and collected by two silicon-based SPADs. The coincidence histogram yields the second-order correlation function $g^{(2)}(\tau)$; a lower $g^{(2)}(0)$ value indicates higher single-photon purity while $g^{(2)}(0)<0.5$ is generally recognized as the criterion for a single-photon emitter \cite{lin2017electrically,yan2022double}.

Figure 3(c) shows the photon second-order correlation result for a bare CQD, exhibiting a pronounced antibunching dip with $g^{(2)}(0)\approx0.119$, consistent with single-photon emission. The experimental data are well fitted by the analytic description of second-order correlation function (bold red line). Figure 3(d) presents the corresponding measurement for a single CQD embedded within a polystyrene nanoparticle (PSNP), yielding $g^{(2)}(0)\approx0.121$. The close similarity of these values demonstrates that the single-photon emission properties of CQDs are largely preserved after polystyrene encapsulation, indicating minimal influence from the polymer environment and synthesis process.

\begin{center}
  \includegraphics[width=\textwidth]{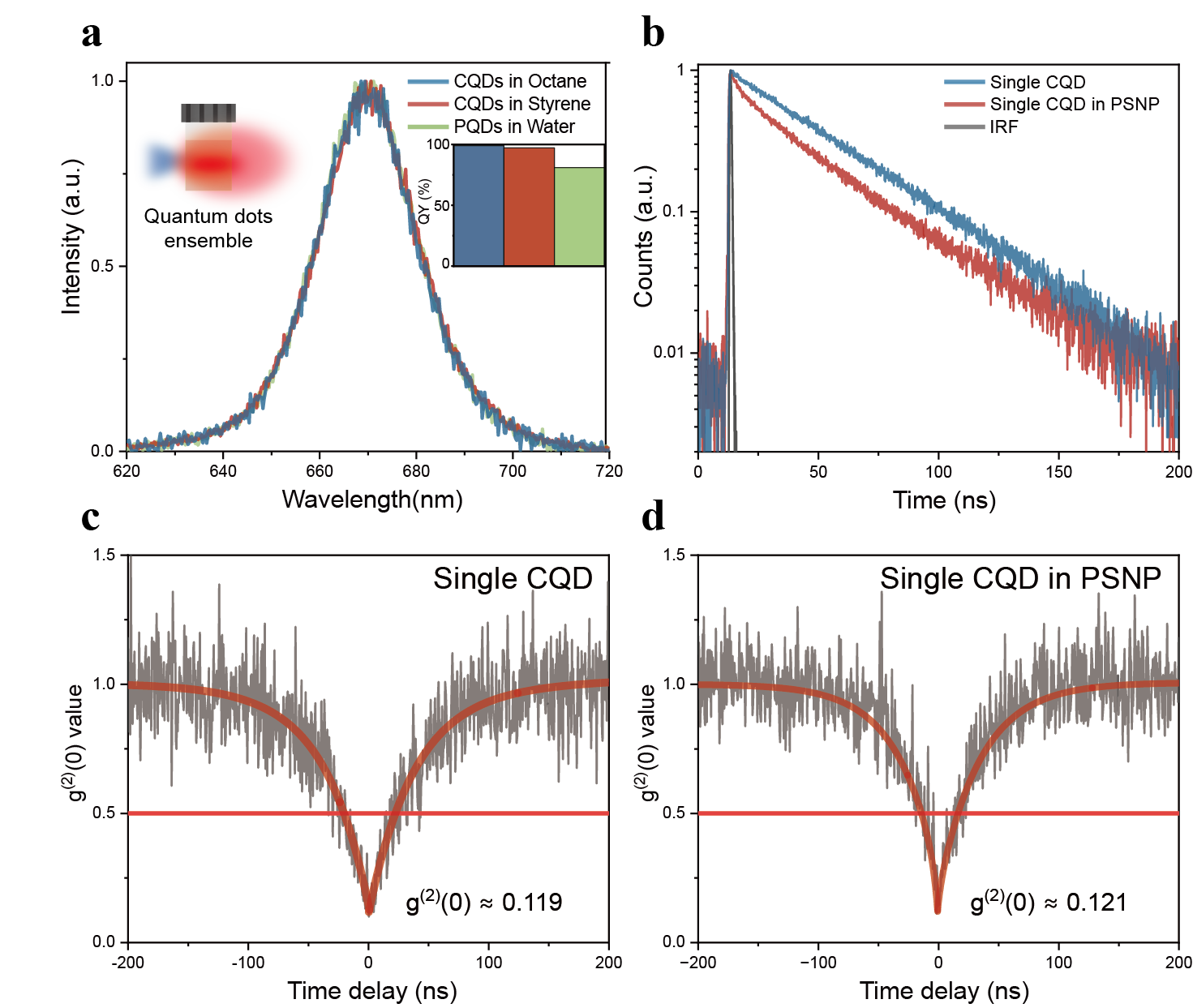}
\end{center}
\vspace{0pt}
\noindent \textbf{Figure 3: Optical characterization of CQDs and PQDs.} (a) PL spectra of PQDs in water and bare CQDs in oil, showing a strong spectral overlap; inset: quantum yields in different environments, for PQDs averaging above 80\%. (b) TRPL spectra of individual CQD and PQD nanoparitcles. (c) Photon correlation of a single CQD, showing antibunching with $g^{(2)}(0)\approx 0.119$. (d) Photon correlation of a PQD with a single emitter inside yields $g^{(2)}(0)\approx 0.121$, confirming the preservation of single-photon characteristics after encapsulation process.
\vspace{15pt} 

\clearpage
\subsection{Collective Emission from Colloidal Quantum Dots}

\begin{center}
  \includegraphics[width=\textwidth]{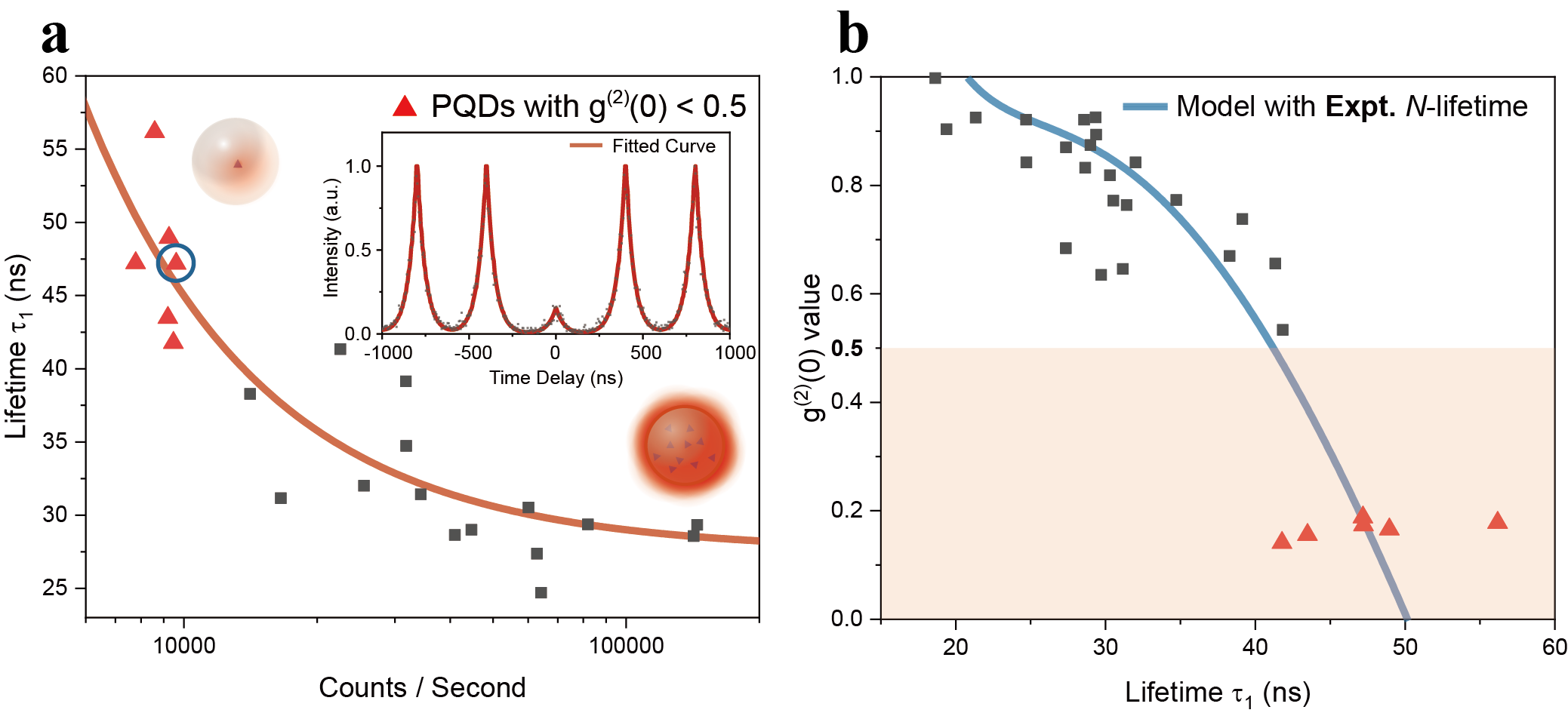}
\end{center}
\vspace{0pt}
\noindent \textbf{Figure 4: Collective emission in single PQD nanospheres.} (a) Photoluminescence lifetime analysis of PQDs, showing that photon lifetime of PQDs decreases with increasing brightness. Red triangles indicate nanospheres with $g^{(2)}(0) < 0.5$. The inset presents the second-order correlation measurement of the dot marked by a blue circle. (b) Correlation between the photoluminescence lifetime of PQDs and their $g^{(2)}(0)$ values, demonstrating that the lifetime decreases as $g^{(2)}(0)$ increases.
\vspace{15pt} 

The optical properties of CQDs in polystyrene encapsulation are well preserved at both ensemble and single-particle levels. During the polymerization process, the weak affinity between CQD surface ligands and surfactant molecules prevents the formation of stable chemical bonds. This leads to a random distribution of CQDs in the oil phase\cite{joumaa2006synthesis}. As a result, an individual polystyrene nanosphere encapsulates a number of CQDs ranging from a single QD to more than ten. In optical characterization, nanospheres containing multiple CQDs is distinguished from single CQDs by their stronger photoluminescence brightness and higher photon count rates recorded by SPADs. Furthermore, a $g^{(2)}(0)$ value greater than 0.5 provides quantitative evidence for the presence of multiple emitters within a single nanoparticle.

Figure 4 presents time-correlated characterization of the PQDs with single and multiple CQDs encapsulated in individual polystyrene nanospheres. To avoid photon pile-up effects at high count rates, which could distort photoluminescence lifetime measurements, neutral density filters were used to attenuate the photon flux and regulate the detection rate of SPADs. The time-domain characterization involves the emission dynamics of CQDs at the room-temperature where CQDs exhibit complex behavior arising from both radiative and non-radiative pathways, which lead to multi-exponential decay profiles. In practice, bi-exponential fitting is commonly used, typically resolving two components with distinct decay rates $\gamma_{1}$ and $\gamma_{2}$\cite{Aubret2016}. The smaller $\gamma_{1}$ corresponds to a bright ``on'' state, in which neutral QDs undergo predominantly radiative recombination and yield strong photon emission \cite{schlegel2002fluorescence}. In contrast, the larger $\gamma_{2}$ is attributed to a dim or ``off'' state, where photo-charging or ionization opens non-radiative channels that accelerate decay rate and suppress radiative emission\cite{cordones2011direct}. Multi-exciton states, though presented only in small fractions, also contribute to the fast component $\gamma_{2}$ \cite{narvaez2006excitonic}. Since it is often difficult to accurately resolve all lifetime channels from experimental TRPL data, the analyses below primarily focus on $\gamma_{1}$, which represents neutral single-exciton recombination. The corresponding lifetime is given as $\tau_1=1/\gamma_1$. For multiple emitters confined within a subwavelength volume, the collective lifetime $\tau_c$ is identified with the measured $\tau_1$.

Figure 4(a) shows the relationship between the collected photoluminescence count rate and lifetime $\tau_1$ of individual PQDs. A clear negative correlation is observed. As the emission count rate increases, the corresponding lifetime gradually decreases. This trend reflects the presence of multiple CQDs within a single nanosphere, where collective emissions lead to shortened lifetimes \cite{liu2020fourier,raino2018superfluorescence}. In contrast, nanospheres containing only a single CQD display lower emission intensities and longer lifetimes. Nanospheres with $g^{(2)}(0) < 0.5$ are marked by red triangles in the figure. The inset includes a representative antibunching trace corresponding to the blue circled dot. 

In Figure 4(b), $g^{(2)}(0)$ values exhibit a negative correlation with the photoluminescence lifetimes $\tau_1$ of individual PQD nanospheres. Shorter lifetimes are associated with larger $g^{(2)}(0)$ values, reflecting attenuated photon anti-bunching and the involvement of multiple emitters within a single nanosphere. This observation is consistent with the trend in Figure 4(a) and provides complementary evidence that quantum collective behavior gradually dominates when multiple CQDs are confined in close proximity. For the single emitter regime near the lower right corner of the figure, measured $g^{(2)}(0)$ values exhibit negligible variation, indicating that single-photon statistics remain robust. Apart from this, the observed deviation of fluorescence lifetime is attributed to local fluctuations in the dielectric environment introduced during the capping of the CQDs at different spatial locations. The average fluorescence lifetime of single emitters is selected as the reference value. The blue curve represents a theoretical reconstruction based on Equation (4), accounting for the functional dependence of $g^{(2)}(0)$ on the collective lifetime $\tau_1$. By incorporating an experimentally-derived relationship into the model, the emitter number $N$ is formulated as the functions of $g^{(2)}(0)$ and $\tau_1$. A detailed discussion regarding this parameterization is provided in the following section. 

Experimentally, observing photon bunching ($g^{(2)}(0) > 1$) in such colloidal systems is challenging due to two inherent suppression mechanisms. First, under weak-coupling conditions, the variance of the collective decay rates is relatively small. This renders the second term in Equation (3) negative. Second, the intrinsic inhomogeneity of CQDs contributes to the third term. Since this variance appears with a negative sign in the equation, it further reduces the correlation value. Consequently, the combined influence of weak cooperative coupling and emitter inhomogeneity ensures that $g^{(2)}(0)$ consistently remains below 1.

To ensure the reliability of the analyses, we further examine the optical stability of single PQDs\cite{Negele2013}. As shown in \textit{Supplementary Information}, the photon trace of an individual PQD demonstrates stable emission over extended periods. While the majority of PQDs exhibits high stability, a minor fraction shows relatively frequent incidence of optical blinkings, which influence the measured photon lifetime. These cases can be clearly identified and distinguished from a long stable period of the population. Therefore, the presence of such blinking does not compromise the validity of our conclusions on collective behavior and number-resolving studies.

\subsection{Numbering Colloidal Quantum Dots in a Subwavelength Volume}

\begin{center}
  \includegraphics[width=1.0\textwidth]{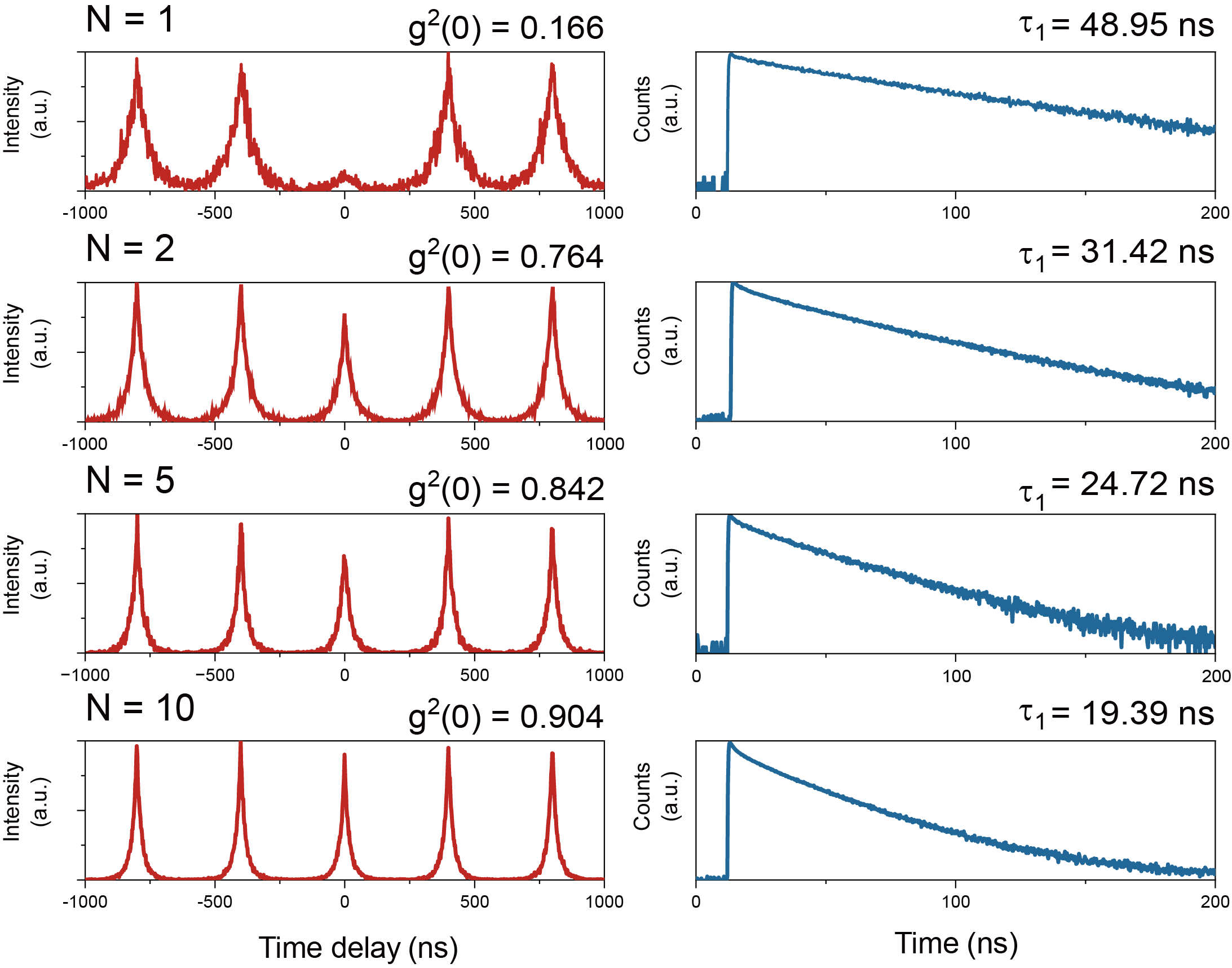}
\end{center}
\vspace{0pt}
\noindent \textbf{Figure 5: Representative PQDs with time-correlated optical responses.} $g^{2}(0)$ and TRPL spectra of four representative PQDs. With increasing emitter number $N$, fluorescence lifetimes become shorter and $g^{(2)}(0)$ values increase. This evolution highlights the transition from single- to few-emitter dynamics and the acceleration of collective decay within nanospheres.
\vspace{15pt}

After confirming the quantum collective behavior of PQDs, we further proceed to quantify the number of active quantum emitters in each nanoparticle. The calculation is performed according to equation (4) using experimentally obtained time-domain characteristics. Depending on the measured photon correlation, $g^{(2)}(0)$ values with single emitters distinguish themselves from the multi-emitter cases. For multi-emitter systems where $g^{(2)}(0) > 0.5$, the emitter number $N$ is determined as the nearest integer to the non-integer solution which is obtained from the model. Figure 5 presents four representative PQDs with distinct optical characteristics. The corresponding effective emitter numbers $N$ are found to be 1, 2, 5, and 10, with extracted lifetimes $\tau_1$ of 48.95 ns, 31.42 ns, 24.72 ns, and 19.39 ns, respectively, showing a systematic reduction in collective lifetime with increasing $N$ and $g^{(2)}(0)$. These results suggest a progressive acceleration in radiative decay as more emitters are coupled within the subwavelength region. \cite{bradac2017room}.

From a series of quantum optical measurements on PQDs, the emitter number $N$ is determined by the analytic extraction for each nanospheres. Figure 6 summarizes how the extracted values of emitter number $N$ depend on fluorescence lifetimes and $g^{(2)}(0)$ values. Experimentally, in Figure 6(a), the fitted $N$-lifetime relation follows a scaling law of $\tau_1 \propto 1/N$, consistent with the prediction of the quantum Dicke model \cite{raino2020superradiant}. This dependence is well fitted by a relation, $\tau_1 = b + a/N$ (blue curve). The constant offset $b$ represents a non-zero floor because the effective decay rate is physically bounded by intrinsic dephasing mechanisms and ambient thermal fluctuations. Simultaneously, the $g^{(2)}(0)$ values (rectangles) show an observed monotonic increase with $N$. Nanospheres capsulizing single emitters show strong antibunching, while nanospheres containing multiple CQDs gradually approach the classical limit of $g^{(2)}(0)=1$. By substituting the $N$-lifetime relation into equation (4), we obtain
\begin{equation}
  g^{(2)}(0) 
  = 1 - \frac{1}{N}
  + \frac{(N-1)\,\bar{\tau}_0^{\,2}}
  {N\!\left(a+bN\right)^2}.
\end{equation}
This equation provides a direct quantitative link between the photon-correlation parameter $g^{(2)}(0)$ and the emitter number $N$. The derived function is then successfully used to describe the experimentally observed $N$-$g^{(2)}(0)$ correlation (red curve). The two relations and their visualizations in Figure 6(a) clearly indicate the analytic dependence of the emitter number $N$ on the values of $g^{(2)}(0)$ and $\tau_1$, providing a solid ground for the strict numbering of quantum emitters based on optical inspections. 

Figure 6(b) presents a three-dimensional surface mapping of the theoretical relationship between collective lifetime $\tau_1$, photon correlation $g^{(2)}(0)$, and emitter number $N$ in the form of discrete integers. In this map, distinct color zones correspond to specific emitter numbers ranging from $N=2$ to 10 within the multi-emitter regime. In most regions, the surface is monotonic, corresponding to a unique $N$ and allowing for its direct identification. However, within limited parameter spaces, colored strips can go across each other, numerically corresponding to multiple-root solutions. To ensure accuracy, the solution space is further constrained by the fluorescence intensity of the PQD, adopting values consistent with weak-coupling conditions. The red curve in the plot represents the calculation of equation (5) based on experimentally derived parameters, e.g. $N$-lifetime relation. While the trajectory in red aligns well with the theoretical surface, it provides a global characterization of the entire dataset rather than a precise description of every individual data point. The blue curve is a projection of the red trajectory extending toward higher $N$ values, corresponding to the curve shown in Figure 4(b). The projection curve is truncated as $g^{(2)}(0)$ approaches 1. It is a consequence of the baseline floor in the fitted $N$-lifetime relationship. In the regime of much smaller $\tau_1$, the emitter count $N$ loses its physical meaning. The constrained model curve serves primarily to illustrate a common trend rather than a rigorous tool for strict emitter number resolving. Compared with parameter fitting, the 3D surface representation of this methodology provides a robust and intuitive framework for the deterministic $N$-s based on experimentally measured $\tau_1$ and $g^{(2)}(0)$. This approach highlights the evolution from single-emitter antibunching to multi-emitter collective emission behaviors.

\begin{center}
  \includegraphics[width=\textwidth]{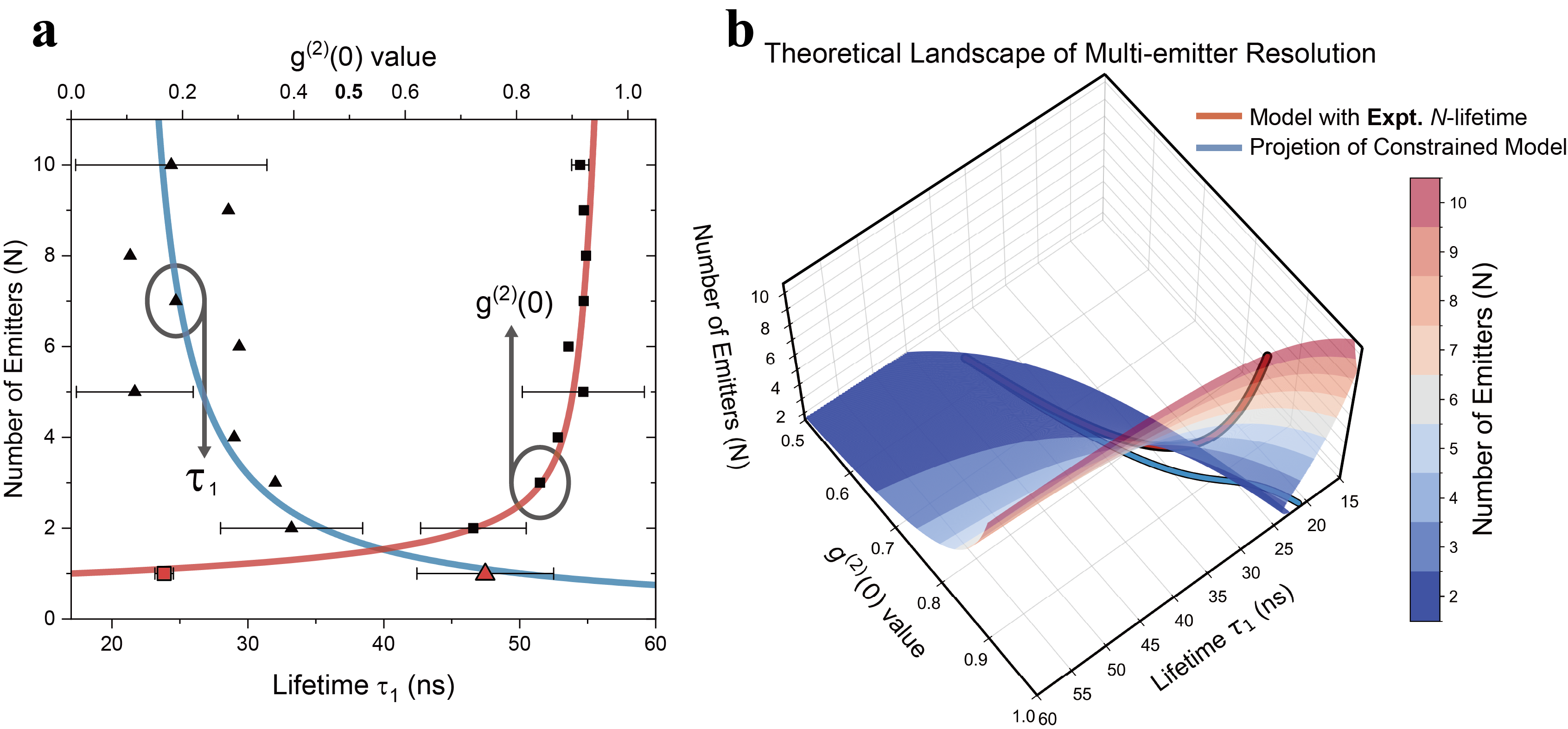}
\end{center}
\vspace{0pt}
\noindent \textbf{Figure 6: Evolution of emission dynamics and photon statistics dependent on emitter number $N$.} (a) $g^{(2)}(0)$ values increase with $N$, shifting from clear antibunching in the single-emitter regime and converging toward the classical limit as the emitter number increases. Fluorescence lifetime $\tau_1$ decreases with increasing $N$, and the fitted $N$–lifetime curve reveals a correlation of $\tau_1 \propto 1/N$, consistent with collective emission dynamics. (b) Three-dimensional visualization illustrating the relationship between fluorescence lifetime and photon correlation, serving as a map for resolved emitter number identification in the multi-emitter regime. The overlaid red curve depicts the theoretical model constrained by experiemental parameters representing the statistical trend, while the blue curve serves as its projection.
\vspace{20pt} 

In general, Figure 6 demonstrates that both photon correlation and decay dynamics provide consistent signatures of quantum collective emission in PQDs. The simultaneous evolution of $g^{(2)}(0)$ and $\tau_1$ with $N$ establishes a framework for the quantum optical resolution of emitters in a spatially confined subwavelength volume. Although the current experiment only resolves the emitter number up to $N=10$, this quantum optical framework remains valid for larger $N$-s. The upper limit of this methodology is governed primarily by the temporal resolution of the detection system required to distinguish the converging values of $g^{(2)}(0)$ and $\tau_1$.

\begin{center}
  \includegraphics[width=\textwidth]{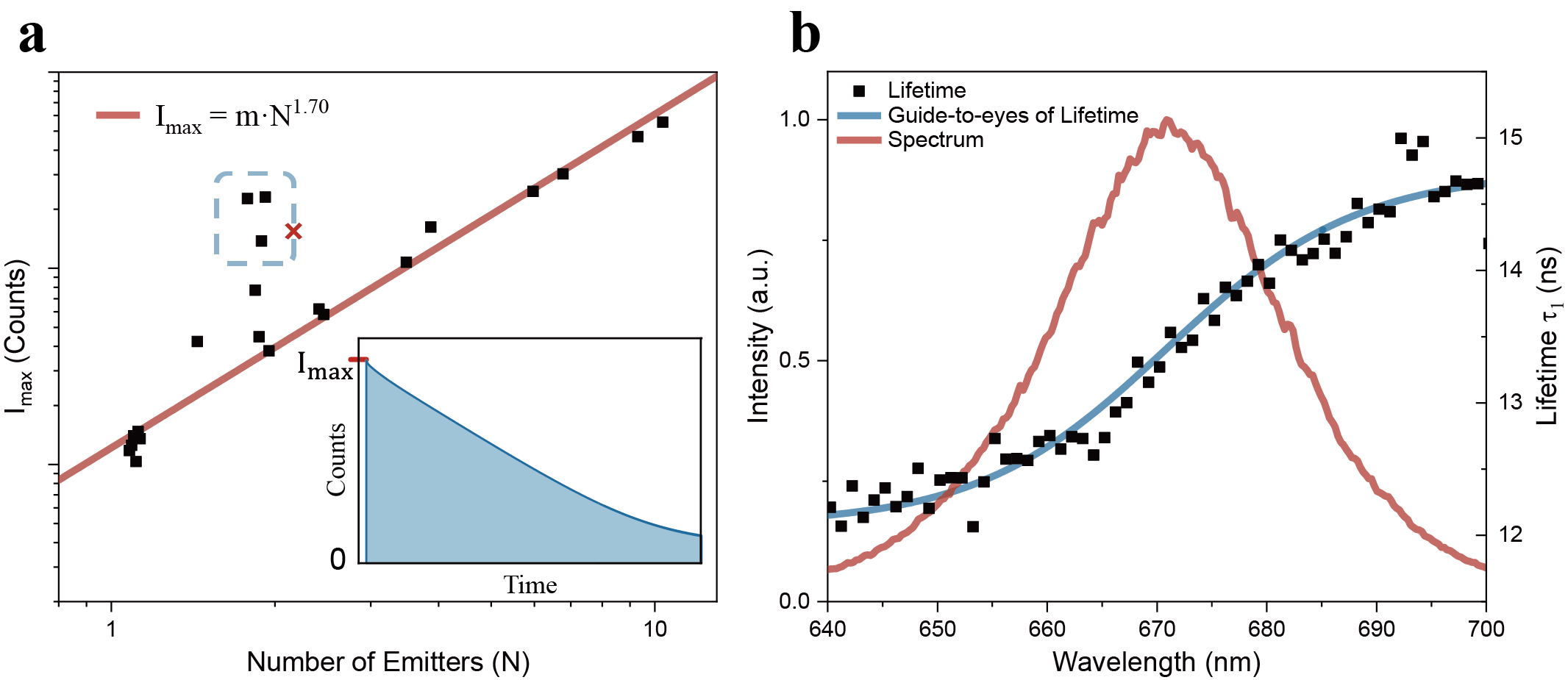}	
\end{center}
\vspace{0pt}
\noindent \textbf{Figure 7: Supplementary experimental evidence for collective emission from multiple emitters within a single polystyrene capsule.} (a) Dependence of the maximum instantaneous fluorescence intensity of PQDs on the emitter number, $N$. The data are fitted to a superlinear power law. The observed superlinear growth (exponent $1.70>1$) indicates coupling among the emitters, a hallmark feature of cooperative emission. The inset illustrates the determination of $I_{max}$ from the TRPL decay profile. (b) Wavelength dependence of photoluminescence lifetime for a multi-emitter PQD nanoparticle. The maximum variation of the lifetime is observed near the peak of the emission wavelength.\vspace{15pt}

As stated above, core-shelled and polystyrene capsulized CQDs provide unique quantum few-body systems, which effectively suppresses direct charge transfer between artificial atoms and guarantees mutual coupling of quantum emitters through coherent optical field. To support our claim that collective emission is from the CQD ensemble uniformly coupled to an external coherent field, we further provide evidence on spontaneous supperradiance from individual PQDs. In Dicke's description, superradiance emission emerges with accelerated lifetime that is inversely proportional to the number of coupled emitters $N$. Based on the law of energy conservation, the peak intensity of time-resolved photoluminescence depends proportionally on $N^2$ \cite{raino2020superradiant}. In the temporal evolution of fluorescence measured from CQD emitters over the same integration time, the maximum photon counts is recorded as the instantaneous peak intensity $I_{max}$. As shown in Figure 7(a), $I_{max}$ exhibits a superlinear dependence on the emitter number $N$, characterized by a power-law exponent of $1.70$. This scaling behavior confirms that the emission intensity grows faster than the linear accumulation of individual emitters, providing direct evidence for collective quantum few-body effects. In the figure, there are a few failure data points with large deviations, highlighted by a dashed blue box, which could be attributed to the complex spatial distribution of emitters. The wavelength dependence of PL lifetimes of a multi-emitter PQD is presented in Figure 7(b). The wavelength dependence of PL lifetime is collected with a spectrum resolution of 1 nm. Consistent with the well-known spectral characteristics of QDs, the photon decay time increases monotonically from the short-wavelength side to the long-wavelength side \cite{malik2001time,patton2003trion}. The guide-to-eyes curve reveals that the most significant variation in PL lifetime coincides with the peak of the emission wavelength \cite{Scheibner2007}. This observation suggests that the main emission peak is largely attributed to coorperative modes rather than independent exciton transitions, which further supports the collective nature of the system.

\section{Conclusions}  

In conclusion, we have experimentally demonstrated a non-invasive method for the number resolution of CQDs confined within a subwavelength volume. By encapsulating CQDs into polystyrene nanospheres, a quantum few-body system based on artificial atoms is created to effectively meet the conditions of Dicke's model on spontaneous superradiance. The feasibility of this approach relies on the unique optical nature of CQDs, especially their non-polarized dipole orientation and homogeneously broadened emission, which ensures uniform mutual coupling through coherent optical field. This methodology guarantees the effective resolution of an emitter number up to ten, showing clear transition from single-emitter to multi-emitter dynamics which follows the scaling law. 

In previous work on solid-state artificial atoms, due to the spreading normal distribution of atomic transition, the strict compliance of the quantum Dicke model has been compromised, placing significant difficulties on the application side of the superradiance phenomenon beyond physics studies. The current work bridges the gap between macroscopic optical measurements and the precise resolution of microscopic quantum quantities for artificial atoms. The polystyrene encapsulated CQD system establishes the desired capability of superraidance phenomenon as a macroscopic tool for quantitative inspection of quantum emitters and thus paves the way for investigating collective light-matter interactions, which provide essential components for the development of quantum photonic technologies and the nano-manipulation for bio-inspections based on fluorescent nanocrystals.

\section{Methods}

\subsection{Chemicals}
\label{startsample}
Core-shelled CQDs (CdSe/8CdS/9ZnS nanocrystals, Najingtek), Sodium bicarbonate ($\rm NaHCO_3$, Aladdin, 99.7\%), Potassium persulfate (KPS, Sigma Aldrich, 99\%), n-Hexadecane ($\rm {C_{16}H_{32}}$, Sigma Aldrich, 99\%), Styrene ($\rm {C_{8}H_{8}}$, Energy Chemical, 99.5\%) and Sodium dodecyl sulfate (SDS, Aladdin, 92.5\%) were used in the polystyrene encapsulation synthesis process.

\subsection{Micro-photoluminescence}
Optical characterizations were carried out by a home-built micro-photoluminescence system. During the experiment, both a continuous-wave (CW) laser and a picosecond laser operating near the wavelength of 450 nm were used to excite the CQDs sample. An oil-immersed objective (Olympus, PlanApo N, NA 1.42) was used to collect the fluorescence of a single QD nanoparticle with a magnification of 60$\times$. Another 20$\times$ air objective (Olympus, LUCPlanFL N, NA 0.45) was used to collect fluorescence from the CQD ensemble in different solutions. The emitted photons were collected by two single photon avalanche photodiode detectors (SPAD, SPCM-AQR-14, Perkin Elmer). Fluorescence lifetime decay and photon coincidence were analyzed by a time-correlated single photon counting module (Hydraharp400, Picoquant). All optical experiments were conducted under ambient conditions. For all comparative measurements, experimental conditions and the excitation power were kept constant.

\subsection{Derivation of Coupled Emitter Number $N$}
We begin with a Markovian master equation describing the dynamics of a system of $N$ two-level emitters in free space, after tracing out the electromagnetic field degrees of freedom. The evolution of the atomic density matrix $\rho$ is given by

\begin{equation}
	\dot{\rho} = -\frac{i}{\hbar}[H, \rho] + \sum_{i,j=1}^{N} \frac{\Gamma_{ij}}{2} \left( 2\sigma_j \rho \sigma_i^\dagger - \sigma_i^\dagger \sigma_j \rho - \rho \sigma_i^\dagger \sigma_j \right).
\end{equation}
Where $\sigma_{i} = |g_i\rangle \langle e_i|$ is the lowering operator for the $i$-th emitter, $\sigma_{i}^{\dagger} = |e_i\rangle \langle g_i|$ is the corresponding raising operator, and \( \Gamma_{ij} \) describes the collective decay rate of the dissipative interaction between emitters \( i \) and \( j \). We now specify the Hamiltonian $H$
\begin{equation}
H = \hbar \sum_{i} \omega_0 \sigma_i^{\dagger} \sigma_i + \hbar \sum_{i \ne j} J_{ij} \sigma_i^{\dagger} \sigma_j.
\end{equation}
Here, $J_{ij}$ represents the coherent coupling strength between emitters $i$ and $j$. Physically, the coherent coupling term \( J_{ij} \) describes the lossless exchange of energy between emitters via virtual photon exchange. In contrast, the collective decay rate \( \Gamma_{ij} \) accounts for irreversible radiative losses, governing dissipative collective dynamics such as superradiance and subradiance. The interaction parameters $J_{ij}$ and
$\Gamma_{ij}$ are related to the electromagnetic Green's function
\begin{equation}
J_{ij} - i \frac{\Gamma_{ij}}{2} = -\frac{\mu_0 \omega_0^2}{\hbar} \, \mathbf{d}^* \cdot \mathbf{G}_0(\mathbf{r}_i, \mathbf{r}_j; \omega_0) \cdot \mathbf{d},
\end{equation}
where $\mathbf{d}$ is the transition dipole moment of the emitters. The Lindblad dissipator in the local basis is given by
\begin{equation}
	\mathcal{D}[\rho] = \sum_{i,j=1}^{N} \frac{\Gamma_{ij}}{2} \left( 2\sigma_j \rho \sigma_i^\dagger - \sigma_i^\dagger \sigma_j \rho - \rho \sigma_i^\dagger \sigma_j \right).
\end{equation}
While the dissipator is written in terms of local operators, the non-diagonal structure of $\Gamma_{ij}$ reflects interactions between emitters mediated by shared environment. As a result, radiative decay does not occur independently in the ensemble, but through collective decay channels in which multiple emitters contribute to photon emission. To make this structure more explicit, the Hermitian matrix $\mathbf{\Gamma}$ is diagonalized, yielding eigenvalues $\Gamma_\nu$ and orthonormal eigenvectors $u_n^{(\nu)}$ which define collective jump operators.
\begin{equation}
	L_\nu= \sum_{n=1}^{N} u_n^{(\nu)} \sigma_n,
\end{equation}
which acts as the effective lowering operator for the $\nu$-th collective mode (eigenmode) of the system. This transformation decouples the system dynamics, allowing us to rewrite the dissipator as a sum over independent collective modes
\begin{equation}
	\mathcal{D}[\rho] = \sum_{\nu=1}^{N} \frac{\Gamma_\nu}{2} \left( 2L_\nu \rho L_\nu^\dagger - L_\nu^\dagger L_\nu \rho - \rho L_\nu^\dagger L_\nu \right),
\end{equation}
where each decay channel is characterized by the rate $\Gamma_{\nu}$. This form makes the emergence of collective emission behavior more intuitive. The total emission rate of the CQD ensemble writes
\begin{equation}
	R = \sum_{\nu=1}^{N} \Gamma_{\nu} \left\langle L_{\nu}^\dagger L_{\nu} \right\rangle.
\end{equation}
To derive the explicit form of the second-order correlation function, we consider the photon statistics of the emitter ensemble. As discussed in Section 2.1, the homogeneously broadened emission of CQDs allows us to treat the emitters as identical. Consequently, the inhomogeneity term vanishes. Furthermore, on the assumption of a fully excited state where all emitters are initially populated, the expectation values of the operators can be evaluated directly. The simplified second-order correlation function is then given as\cite{masson2022universality,patti2021controlling}
\begin{equation}
	g^{(2)}(0) = \frac{\sum_{\nu, \mu = 1}^{N} \Gamma_{\nu} \Gamma_{\mu}\left\langle L_{\nu}^\dagger L_{\mu}^\dagger L_{\mu} L_{\nu} \right\rangle}{\left(\sum_{\nu=1}^{N} \Gamma_{\nu}\left\langle L_{\nu}^\dagger L_{\nu} \right\rangle\right)^2} 
	= 1+\frac{1}{N} \left[\mathrm{Var} \left( \frac{\{\Gamma_\nu\}}{\bar{\Gamma}_0} \right)-1\right].
\end{equation}
Since resolving individual lifetimes for all channels is experimentally challenging, we adopt a dominant radiative channel model as a physical simplification. This is consistent with Dicke’s assumption at small emitter separations, where most jump operators correspond to dark channels, while only a few remain bright with large emission rates. Specifically, we assume that only the main emission channel remains bright with a large decay rate $\Gamma_c$
\begin{equation}
	\Gamma_\nu = \left\{
	\begin{array}{ll}
		\Gamma_c, & \nu = 1 \\
		0, & \nu > 1
	\end{array}
	\right..
    \end{equation}
To account for fluctuations in decay rates within the ensemble under this simplification, we apply a variance-based statistical framework. This framework provides a more robust description of the coupling dynamics. Based on this distribution, the variance is calculated as:
\begin{equation}
	N\cdot \mathrm{Var} \left( \frac{\{\Gamma_\nu\}}{\bar{\Gamma}_0} \right) 
	= (N - 1) \left( \frac{\Gamma_c}{N \bar{\Gamma}_0} \right)^2 
	+ \left( \frac{\Gamma_c}{\bar{\Gamma}_0} - \frac{\Gamma_c}{N \bar{\Gamma}_0} \right)^2 
	= \frac{(N - 1)\Gamma_c^2}{N \bar{\Gamma}_0^2}.
\end{equation}
Upon this assumption, the expression for $g^{(2)}(0)$ goes to,

\begin{equation}
	g^{(2)}(0)= 1+ \frac{1}{N}\left[\frac{(N - 1)\Gamma_c^2}{N^2 \bar{\Gamma}_0^2}-1\right].
\end{equation}
By identifying the neutral single-exciton recombination rate, ${\gamma}_1$, with the collective decay rate, ${\Gamma}_c$, we obtain a simplified expression. This form explicitly reveals how the second-order correlation at zero-time delay is governed by the interplay between the quantum emitter number and the ratio of the collective decay rate to the average single-particle decay rate, $\bar{\Gamma}_0$.

\section{Data availability}
The data that support the findings of this study are available from the corresponding author upon reasonable request.

\section{Acknowledgements}
Z.N. would like to thank Prof. Xiao-Gang Peng and all students in his research group (Jia-Kuan Zhang, Xiong-Lin Zhou, Zhe Wang, Xu Cao, et al.) from Department of Chemistry, Zhejiang university for helpful discussions in chemical synthesis.
This work is supported by the Zhejiang Province Leading Geese Plan (2024C01105), the National Future Industry Innovation Mission, the Beijing Natural Science Foundation (L248103), the National Key Research and Development Program of China (2021YEB2800500), and the National Natural Science Foundation of China (61574138, 61974131).

\section{Author contributions}
Z.N., C.J. and X.L. conceived the idea. J.Y., C.J. and Z.N. developed the theoretical model. J.L. and M.J. synthesized the high-quality CQDs. Z.N., X.L., W.F., X.C., Z.S., C.L, F.L., and C.J. contributed to the preparation of the experimental setup. Z.N. conducted the whole experiments. C.J. and X.L. supervised the project. C.J. directed the project. Z.N. and C.J. wrote the manuscript with contributions from all authors.

\section{Competing interests}
The authors declare no competing interests.

\clearpage

\bibliography{Reference/Ref}
\bibliographystyle{Reference/sn-aps1}
\end{document}